# Efficient metallic spintronic emitters of ultrabroadband terahertz radiation

T. Seifert[1], S. Jaiswal[2,3], U. Martens[4], J. Hannegan[5], L. Braun[1], P. Maldonado[6], F. Freimuth[7], A. Kronenberg[2], J. Henrizi[2], I. Radu[8], E. Beaurepaire[9], Y. Mokrousov[7], P.M. Oppeneer[6], M. Jourdan[2], G. Jakob[2], D. Turchinovich[10], L.M. Hayden[5], M. Wolf[1], M. Münzenberg[4], M. Kläui[2], T. Kampfrath[1,*]

1. Department of Physical Chemistry, Fritz Haber Institute of the Max Planck Society, 14195 Berlin, Germany
2. Institute of Physics, Johannes Gutenberg University, 55128 Mainz, Germany
3. Singulus Technologies AG, 63796 Kahl am Main, Germany
4. Institute of Physics, Ernst Moritz Arndt University, 17489 Greifswald, Germany
5. Department of Physics, University of Maryland Baltimore County, Baltimore, MD 21250
6. Department of Physics and Astronomy, Uppsala University, Uppsala, Sweden
7. Peter Grünberg Institute and Institute for Advanced Simulation, Forschungszentrum Jülich and JARA, Jülich, Germany
8. Institute for Optics and Atomic Physics, Technical University Berlin and Helmholtz-Zentrum Berlin für Materialien und Energie, Berlin, Germany
9. Institut de Physique et Chimie des Matériaux de Strasbourg, France
10. Max Planck Institute for Polymer Research, 55128 Mainz, Germany

[*] E-mail: kampfrath@fhi-berlin.mpg.de

**Terahertz electromagnetic radiation is extremely useful for numerous applications such as imaging and spectroscopy. Therefore, it is highly desirable to have an efficient table-top emitter covering the 1-to-30-THz window whilst being driven by a low-cost, low-power femtosecond laser oscillator. So far, all solid-state emitters solely exploit physics related to the electron charge and deliver emission spectra with substantial gaps. Here, we take advantage of the electron spin to realize a conceptually new terahertz source which relies on tailored fundamental spintronic and photonic phenomena in magnetic metal multilayers: ultrafast photo-induced spin currents, the inverse spin-Hall effect and a broadband Fabry-Pérot resonance. Guided by an analytical model, such spintronic route offers unique possibilities for systematic optimization. We find that a 5.8-nm-thick W/CoFeB/Pt trilayer generates ultrashort pulses fully covering the 1-to-30-THz range. Our novel source outperforms laser-oscillator-driven emitters such as ZnTe(110) crystals in terms of bandwidth, terahertz-field amplitude, flexibility, scalability and cost.**

## Introduction

The terahertz (THz) window, loosely defined as the frequency range from 0.3 to 30 THz in the electromagnetic spectrum, is located between the realms of electronics and optics[1,2]. As this region coincides with many fundamental resonances of materials, THz radiation enables very selective spectroscopic insights into all phases of matter with high temporal[3,4] and spatial[5,6,7,8] resolution. Consequently, numerous applications in basic research[3,4], imaging[5] and quality control[8] have emerged.

To fully exploit the potential of THz radiation, energy-efficient and low-cost sources of ultrashort THz pulses are required. Most broadband table-top emitters are driven by femtosecond laser pulses that generate the required THz charge current by appropriately mixing the various optical frequencies[9,10]. Sources made from solids usually consist of semiconducting or insulating structures with naturally or artificially broken inversion symmetry. When the incident photon energy is below the semiconductor band gap, optical rectification causes a charge displacement that follows the intensity envelope of the incident pump pulse[9,10,11,12,13,14,15,16,17]. For above-band-gap excitation, the response is dominated by a photocurrent[18,19,20,21,22,23,24] with a temporally step-like onset and, thus, generally smaller bandwidth than optical rectification[9]. Apart from rare exceptions[14], however, most semiconductors used are polar[1,2,12,13,15,16,17,21,22] and strongly attenuate THz radiation around optical phonon resonances, thereby preventing emission in the so-called Reststrahlen band located between ~1 and 15 THz.

The so far most promising sources covering the full THz window are photocurrents in transient gas plasmas[9,10,25,26,27,28,29]. The downside of this appealing approach is that the underlying ionization process usually requires amplified laser pulses with high threshold energies on the order of 0.1 mJ. Measurable THz waveforms can be obtained with pump-pulse energies down to[29] ~1 µJ, which is, however, still 2 to 3 orders of magnitude larger than what low-cost femtosecond laser oscillators can provide.

Another promising material class for realizing THz sources are metals[30] because they exhibit a pump absorptivity largely independent of wavelength[31], short electron lifetimes of ~10 to 50 fs[32] (implying broadband photocurrents), a featureless THz refractive index[33] (favoring gap-free emission) and a large heat conductivity (for efficient removal of excess heat). In addition, metal thin-film stacks (heterostructures) are well established, simple and cheap to fabricate. Recent works have indeed demonstrated THz emitters based on metal structures[34,35,36,37]. However, the bandwidth did not exceed 3 THz, and THz field amplitudes competitive with those of ZnTe emitters were obtained only in conjunction with amplified laser pulses[36,37]. Consequently, the full potential of metal-based THz emitters is far from being realized.

At this point, it is important to acknowledge that all previously demonstrated THz emitters have taken advantage exclusively of the charge but not the spin of the electron. On the other hand, very recent tremendous progress in the fields of spintronics[38,39,40,41] and femtomagnetism[42,43] has shown that the electron spin offers entirely new possibilities to generate transient currents in metals. In fact, spin-to-charge conversion has been revealed lately as a new pathway to ultrafast photocurrents[44,45].

In this Article, we exploit the unique spintronic and photonic properties of ultrathin metal films to realize a THz emitter driven by ~1 nJ laser pulses from a compact, high-repetition-rate femtosecond laser oscillator. The new source combines various benefits in one device: large bandwidth, large THz-field amplitude at low pump power, easy operation, scalability and low cost. This achievement becomes possible because our widely tunable approach provides access to a large set of spintronic metals and geometrical parameters for optimization.

**Spintronic THz emitter**

**Concept.** Figure 1a illustrates the basic principle of our spintronic THz emitter by means of a bilayer structure consisting of a ferromagnetic (FM) and nonferromagnetic (NM) metal thin film[44]. The FM layer is magnetized in-plane, antiparallel to the $y$-axis. An incident femtosecond laser pulse excites electrons in the metals to states above the Fermi energy, thereby changing their band velocity and scattering rate. Since FM and NM layers have different transport properties, a net current along the $z$-axis is launched. In addition, because the product of density, band velocity and lifetime of spin-up (majority) electrons in FM metals (such as Fe, Co and Ni) is significantly higher than that of the spin-down (minority) electrons[32,46], the $z$-current is strongly spin-polarized[42].

Upon entering the NM layer, spin-orbit interaction deflects spin-up and spin-down electrons in opposite directions[38,39,40,41] by a mean angle $\gamma$ (Fig. 1a). This inverse spin-Hall effect (ISHE) converts the longitudinal ($z$-directed) spin current density $j_s$ into an ultrafast transverse ($x$-directed) charge current density $j_c = \gamma j_s$, thereby acting as a source of THz radiation (Fig. 1a).

**First emission results.** In our experiment, the THz source is excited by femtosecond pulses from a Ti:sapphire laser oscillator (duration 10 fs, center wavelength 800 nm, pulse energy 2.5 nJ, repetition rate 80 MHz). The transient electric field of the emitted THz pulse is measured by electrooptic sampling[1,2,47] in suitable materials (see Methods). We start with bilayers consisting of FM $Co_{20}Fe_{60}B_{20}$ (thickness of 3 nm) capped by either NM Ta or Ir (3 nm) (see Methods). Typical THz electrooptic signals $S(t)$ obtained from these samples magnetized along the $y$-axis (Fig. 1a) are displayed in Fig. 1b.

Consistent with the generation mechanism outlined above[44], the emitted radiation has the following properties (see Supplementary Fig. S4): it is linearly polarized with the electric-field direction perpendicular to the sample magnetization, but it is independent of the pump polarization. The spin current and, thus, THz field are reversed entirely, either when the external saturating magnetic field of 10 mT is reversed or when the two metallic films are grown in reverse order on the substrate. Finally, the THz signal amplitude grows linearly with the pump power (inset of Fig. 1b). Therefore, signal saturation due to pump-induced sample demagnetization is negligible here, consistent with the estimate that ultrafast magnetization quenching is less than 1% at the maximum pump fluence used[44].

Fourier-transforming the time-domain signals $S(t)$ yields the spectral amplitude $|S(\omega)|$ versus frequency $\omega/2\pi$. A typical spectrum is shown in Fig. 1c. It covers the large bandwidth from about 1 to 18 THz. Note that spectral features such as the dip at 8 THz arise from the 50 μm thick GaP electrooptic sensor and not from the emitter[44,47]. By deconvoluting the detector response function from the signal $S(t)$, we obtain the THz electric field $E_{det}(t)$ directly in front of the detection crystal (see

Methods). Strikingly, the field amplitude spectrum $|E_{det}(\omega)|$ (Fig. 1c) is remarkably smooth and extends from 1 to nearly 30 THz full width at 10% amplitude maximum, without any gaps. In addition, the spectral phase of the transient field is flat (Fig. 1c), thereby demonstrating the THz pulse is Fourier-limited.

**Emitter theory.** While its bandwidth is already remarkably large, the $Co_{20}Fe_{60}B_{20}$/Ta bilayer (Fig. 1b) generates a THz peak signal about 2 orders of magnitude smaller than what is obtained from a standard emitter in linear THz spectroscopy[9], a 1 mm thick ZnTe(110) crystal (see Fig. 4a). To boost the emitted THz field, we need to understand the key factors that determine it. The amplitude of the $x$-polarized THz field (Fig. 1a) directly behind the multilayer is given by (see Methods)

$$E(\omega) = Z(\omega)e \int_0^d dz\, \gamma(z) j_s(z, \omega). \qquad (1)$$

According to this generalized Ohm's law, the emitted field $E(\omega)$ equals the total charge current $-e \int dz\, \gamma j_s$ times an impedance $Z(\omega)$ which quantifies how efficiently a current is converted into electromagnetic radiation. Here, $d$ is the film thickness, $-e$ is the electron charge, and $1/Z$ can be interpreted as the effective conductance of a parallel connection of all metal layers shunted by the adjacent substrate and air (see Methods).

Equation (1) readily shows that maximizing $Z$, $\gamma$ and $j_s$ will lead to maximum THz output of the emitter for a given pump power. The numerous sample parameters that can be tuned in such an optimization are the FM/NM materials and the geometry of the heterostructure.

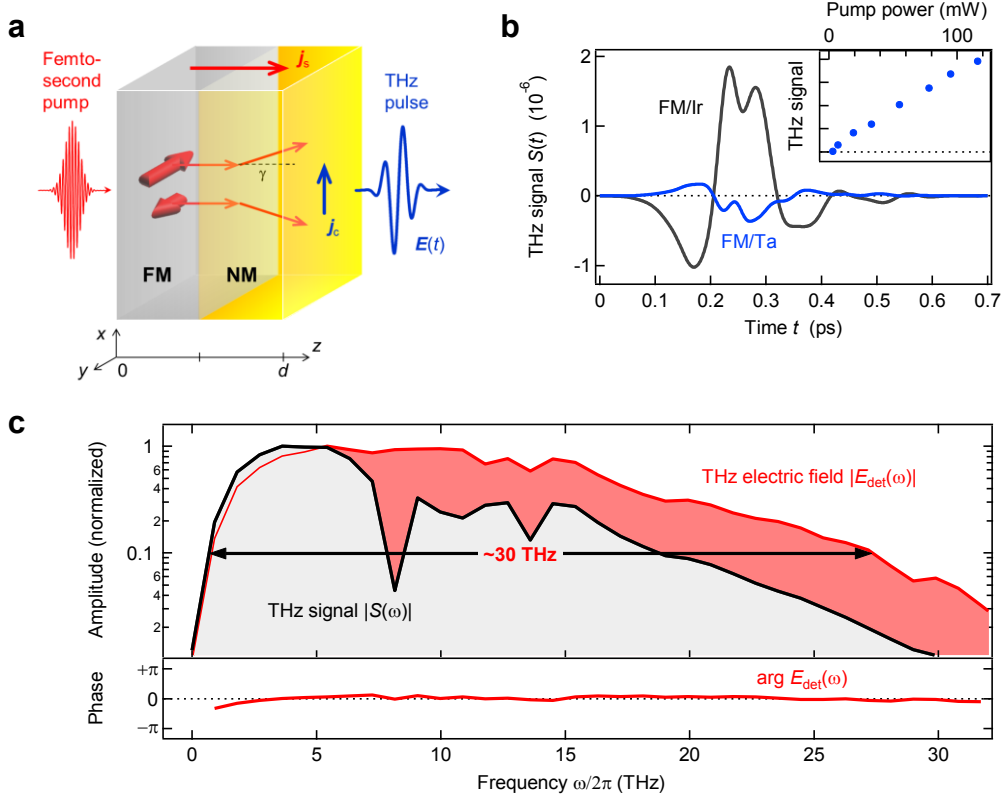

**Figure 1 | Metallic spintronic THz emitter. a**, Principle of operation. A femtosecond laser pulse excites electrons in the metal stack, thereby changing their band velocity and launching a current along the $z$-direction. Since the mobility of spin-up (majority) electrons is significantly higher than that of spin-down (minority) electrons, the $z$-current is spin-polarized. In the NM layer, spin-orbit interaction deflects spin-up and spin-down electrons in opposite directions and transforms the spin current $j_s$ into an ultrafast transverse charge current $j_c = \gamma j_s$, leading to the emission of a THz electromagnetic transient. **b**, Typical electrooptic signal $S(t)$ of THz pulses obtained from photoexcited Ta- and Ir-capped $Co_{20}Fe_{60}B_{20}$ thin films and detected by a 50 μm thick GaP crystal. Inset: THz signal amplitude as a function of the incident pump power. **c**, Fourier spectra of the THz signal $S(t)$ and the extracted transient THz electric field $E_{det}(t)$ incident onto the electrooptic detector. Both spectra are normalized to peak amplitude 1. The double-arrow illustrates the about 30 THz large bandwidth of the emitter. The flat spectral phase attests the THz pulse is Fourier-limited.

**Maximizing the THz output**

**FM/NM materials.** We start with varying the NM material which primarily affects the emitted THz field through the magnitude and sign of the spin-Hall angle $\gamma$ [see Eq. (1)]. Consequently, we in particular consider such metals for which large $\gamma$ values have been reported[39]. Importantly, for all samples studied here, we find emitted THz waveforms and spectra whose shape is very similar to those shown in Figs. 1b and 1c. Therefore, it is sufficient to quantify the strength of the emitted THz field by the root mean square (RMS) of the THz signal $S(t)$. This quantity is displayed in Fig. 2 as a function of the NM metal in $Co_{20}Fe_{60}B_{20}$(3 nm)/NM(3 nm) heterostructures.

Figure 2 clearly shows that the THz field amplitude and polarity depend drastically on the NM material chosen: for instance, Pt delivers a 1 order of magnitude larger amplitude than Ta and Ir. Interestingly, choosing W for the NM layer leads to a comparable magnitude as with Pd or Pt, but with opposite sign. The sign change is consistent with previous works[39] and related to the half-filled d-shell of W yet almost full d-shell of Pt[48]. More generally, we find that the entire trend of THz amplitude versus NM material (Fig. 2) is in good semiquantitative agreement with spin-Hall conductivities measured previously[39] and calculated by us (see Fig. 2 and Methods). This observation provides further evidence for the transport scenario outlined in Fig. 1a and Ref. 44.

In contrast to the NM material, the THz signal amplitude is found to change only relatively little when the FM material $Co_{20}Fe_{60}B_{20}$ is substituted by Fe, Co, Fe-Co alloys or $Ni_{81}Fe_{19}$ (see Supplementary Section S1). In essence, material variation of the FM/NM bilayers shows that the combination $Co_{40}Fe_{40}B_{20}$/Pt provides best THz-emission performance.

**Sample thickness.** In the next optimization step, we vary the stack geometry and measure THz emission as a function of the total sample thickness $d$ while keeping the FM and NM layer approximately equally thick. The experiment reveals a surprising behavior (Fig. 3a): the THz amplitude *increases* with *decreasing* emitter thickness $d$, peaks at $d$=4 nm and falls off rapidly at smaller $d$.

Such behavior is highly counterintuitive and in sharp contrast to most phase-matched frequency conversion schemes such as optical rectification and second-harmonic generation[9]. Indeed, Eq. (1) suggests that the THz amplitude scales with metal thickness $d$. This trend, however, is overcompensated by a remarkable photonic effect: our metal thin film acts as a Fabry-Pérot cavity that resonantly enhances both pump and THz waves. Because the cavity length $d$ is much smaller than all wavelengths involved, all reflection echoes inside the film interfere constructively (Fig. 3b). The shorter the cavity, the more echoes occur before the light wave has decayed, resulting in even more enhancement. Below a critical thickness $d_c$, however, reflection losses at the cavity faces exceed attenuation in the metal bulk. Then, the enhancement of pump and THz electric field saturates at $d<d_c$ and no longer compensates for the shrinking emitter volume. Therefore, the emitted THz amplitude should first grow with decreasing $d$ and, after reaching a maximum, decrease, in agreement with our experimental bilayer data (Fig. 3a).

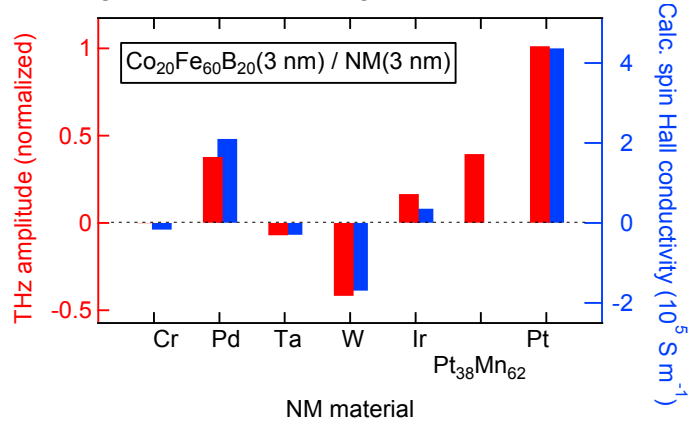

**Figure 2 | Impact of the NM material.** THz signal amplitude (RMS) as a function of the NM material used for the $Co_{20}Fe_{60}B_{20}$(3 nm)/NM(3 nm) stack (red bars). For comparison, *ab initio*-calculated values of the spin-Hall conductivity are also shown (blue bars), except $Pt_{38}Mn_{62}$. A Cr cap layer leads to nearly vanishing THz signal.

A quantitative description of this behavior is provided by Eq. (1) and a simple transport model for $j_s$ (see Methods). The best fit (solid curve in Fig. 3a) is obtained when we assume that the spin-polarized carriers entering the Pt layer randomize their velocity over a length of 1 nm. In essence, Fig. 3a shows that the largest THz signal is provided by the 4 nm thick $Co_{20}Fe_{60}B_{20}$/Pt emitter, and this performance is very close to the absolute maximum predicted by our model.

**Trilayer.** Having identified the best bilayer emitter, we finally tailor the sequence of the spintronic metal layers. Figure 1a suggests that only the forward-propagating half of the photoinduced spin current travels into the NM layer where it is converted into a charge current. To also take advantage of the backward-flowing electrons, we introduce another NM layer on the left-hand side of the FM film (Fig. 3c). We choose W/$Co_{40}Fe_{40}B_{20}$/Pt because W and Pt exhibit the largest spin-Hall angles $\gamma$ yet with opposite sign (Fig. 2). Owing to this unique possibility of spintronic engineering, the spin-Hall currents $j_c$ in the W and Pt layer flow in the same direction, radiate in phase and, thus again, boost the THz amplitude (Fig. 3c).

Indeed, as seen in Fig. 3a, the THz amplitude from each W/$Co_{40}Fe_{40}B_{20}$/Pt trilayer is approximately twice as high as that from a bilayer counterpart with the same total thickness. In particular, the W(2 nm)/$Co_{40}Fe_{40}B_{20}$(1.8 nm)/Pt(2 nm) trilayer delivers a 40% higher THz amplitude than the best bilayer $Co_{20}Fe_{60}B_{20}$(2 nm)/Pt(2 nm), even though the trilayer is about 50% thicker. This result indicates that the conversion of both forward and backward spin currents into THz radiation overcompensates the effect of the larger metal thickness.

Figure 3a concludes our model-guided optimization strategy and identifies the 5.8 nm thick W/$Co_{40}Fe_{40}B_{20}$/Pt trilayer as the best THz emitter out of the comprehensive set of more than 70 heterostructures studied here. With this extensive procedure, we have gone beyond all previous approaches for emitter design and fully exploited the spintronic nature of our THz source. The evolution of our efforts is illustrated by Fig. 4a: the trilayer delivers a more than 2 orders of magnitude larger THz amplitude than $Co_{20}Fe_{60}B_{20}$(10 nm)/Ta(2 nm), which is one of the bilayers we started with.

**Performance test**

To evaluate the performance of our trilayer emitter, we compare it to three state-of-the-art THz sources routinely used to cover the range from about 0.3 to 8 THz: the nonlinear-optical crystals ZnTe(110) and GaP(110)[1,9] and a high-performance photoconductive switch with interdigitated electrodes[22] (see Methods). For all emitters, THz emission is measured under identical conditions, and the THz signal amplitude is found to scale linearly with the pump power. Consequently, comparison of THz spectral amplitudes provides a direct measure of how well each emitter performs at a given frequency. To directly identify spectral emission gaps, we choose a 70 µm thick Lemke/amorphous polycarbonate (LAPC) electrooptic sensor[14] that permits gap-free detection from about 0.3 to 15 THz (see Supplementary Fig. S2).

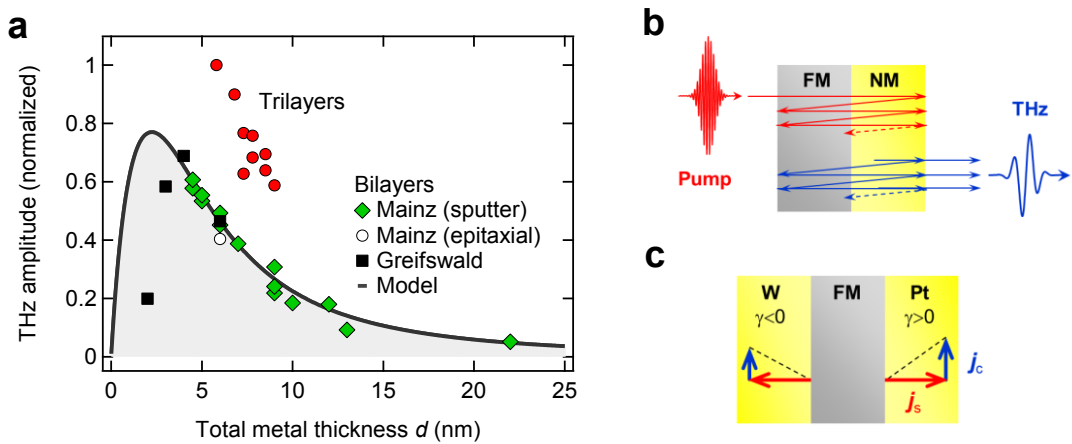

**Figure 3 | Impact of stack geometry on emitter performance. a**, THz signal amplitude (RMS) as a function of the thickness $d$ of the metal stack. Different symbols represent different sample fabrication schemes (see Methods). The solid line is a fit using a transport model (see main text and Methods). **b**, Schematic of the thin-film Fabry-Pérot cavity that enhances both incident pump and emitted THz radiation. **c**, Schematic of the trilayer emitter that converts the backward- and forward-flowing spin current $j_s$ into a unidirectional charge current $j_c$ with approximately equal efficiency.

Figures 4a and 4b display THz waveforms $S(t)$ from all sources and the respective amplitude spectra. The resulting transient THz electric fields (see Methods) are shown in the Supplementary Fig. S1. For ZnTe and GaP, the electrooptic signal $S(t)$ consists of a slowly and fast oscillating part which is, respectively, related to frequencies below and above the Reststrahlen band of these crystals. Strong wave attenuation in the Reststrahlen region[47] leads to considerable gaps from 3 to 10 THz and 7 to 13 THz in the ZnTe and GaP amplitude spectra $|S(\omega)|$, respectively (Fig. 4b). In contrast, the time-domain signal from the spintronic trilayer features a higher peak amplitude, is much shorter (Fig. 4a) and even Fourier-limited (Fig. 1c). Remarkably, the spectrum is gap-free and exceeds the spectral amplitude of the ZnTe and GaP crystals from 2.5 to 14 THz, except small frequency intervals around 6 and 12 THz where GaP and ZnTe, respectively, yield slightly more amplitude. We note that our trilayer emitter is characterized by an effective $\chi^{(2)}$ nonlinear optical coefficient being 5 orders of magnitude larger than that of GaP (see Supplementary Section S2). As revealed above, however, this coefficient is confined to an only ~1 nm thick layer at the FM/NM interfaces.

While the photoconductive switch and the trilayer exhibit comparable THz signal amplitudes in the time domain (Fig. 4a), their performance is complementary in frequency space (Fig. 4b). In the case exceptionally high amplitudes are required below 3 THz, the photoconductive switch is the source of choice. In contrast, the spintronic emitter provides more amplitude above 3 THz and exhibits a much wider bandwidth from 1 to 30 THz without gap (see Supplementary Fig. S1).

We routinely use our trilayer emitter to measure ultrabroadband THz transmission spectra. As an example, Fig. 4c displays amplitude and phase of the complex transmission of a 7.5 µm thick polytetrafluoroethylene (PTFE, Teflon®) sample, obtained by using the trilayer emitter and a 10 µm thick ZnTe(110) electrooptic sensor[47]. Resonant features around 6, 15 and 18 THz are found, in excellent agreement with previous studies[49] using gas-plasma THz emitters which were, however, driven by 5 orders of magnitude more intense pump pulses. We finally note that such broadband and gapless THz spectroscopy would not be possible at all with standard solid-state emitters.

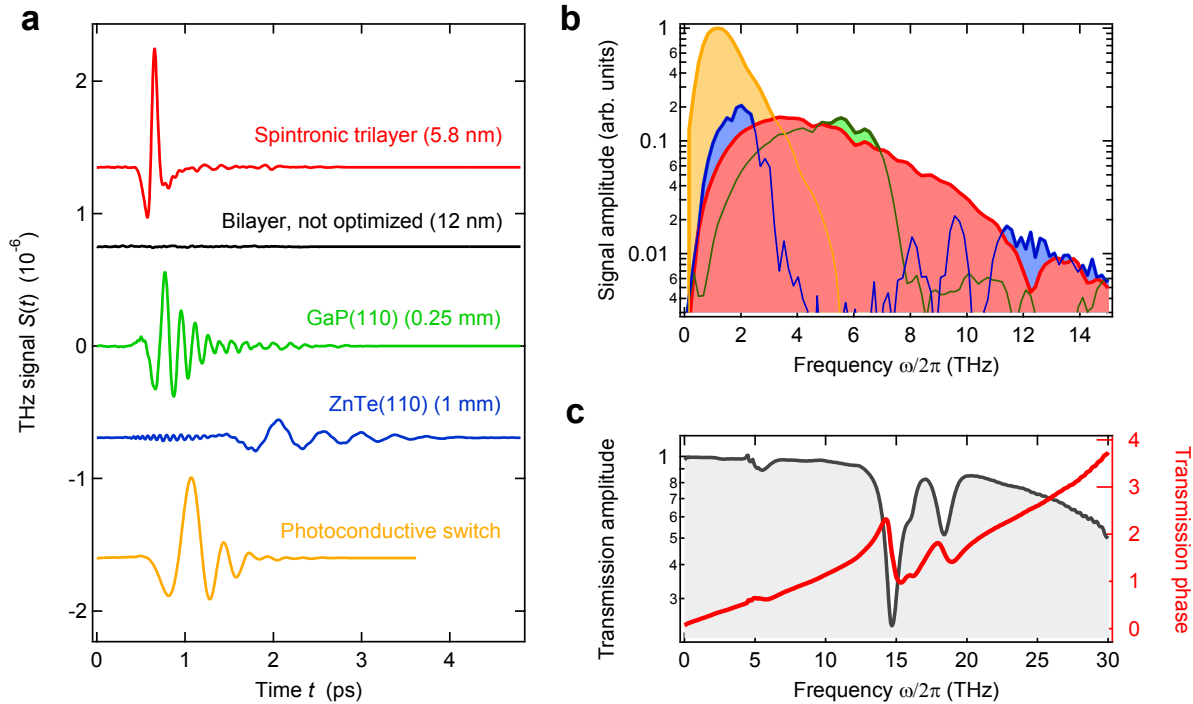

**Figure 4 | Spintronic emitter performance and spectroscopic application. a**, THz signal waveforms and **b**, resulting amplitude spectra of the spintronic trilayer emitter in comparison to standard THz emitters as measured with a 70 µm thick LAPC electrooptic sensor. Measurements are performed under identical conditions such that the output of all emitters can directly be compared. Deconvolution of the detector response function yields the transient THz electric fields which are shown in the Supplementary Fig. S1. **c**, Spectral amplitude and phase of the THz transmission of a 7.5 µm thick PTFE thread-seal tape measured with our spintronic emitter using a 10 µm thick ZnTe electrooptic sensor.

**Conclusion**

We have developed a conceptually new, high-performance and versatile THz source for broadband linear THz spectroscopy based on optically driven spin currents in ultrathin magnetic metal heterostructures. Our approach unifies the benefits of different emitter types in one device: it approaches the bandwidth of much more expensive gas-plasma spectroscopy systems and delivers short, Fourier-limited pulses covering the full range from 1 to 30 THz without gap. As with often-used optical-rectification crystals, our heterostructure is robust, passive, easy-to-use (in transmission mode under normal incidence) and driven by a low-cost, compact femtosecond laser oscillator. At the same time, the THz field amplitude emitted exceeds that of standard emitters such as ZnTe, GaP and a biased photoconductive switch. Similar to such switches, direction and amplitude of the emitted THz field can easily be modulated by applying an oscillating magnetic field with small amplitude below 10 mT. In addition to these benefits, the broadband optical absorption of metals implies that the spintronic THz source can be driven by any laser oscillator, virtually independent of its output wavelength.

We emphasize that fabrication of our emitter is inexpensive, straightforward and scalable, without involving any lithography steps. Fabrication costs are dominated by the substrate price, and we are able to deposit homogeneous layers on substrates with diameters as large as 20 cm (see Methods). Preliminary tests show that driving such large-area trilayers with intense laser pulses easily yields THz pulses with peak fields of several 100 kV cm$^{-1}$ which exceed those obtained with more strongly pumped large-area ZnTe emitters[50]. Therefore, spintronic THz sources exhibit a high potential for enabling nonlinear-optical studies[51,52] in the difficult-to-access region between 5 and 10 THz. More generally, our results highlight metallic magnetic multilayers as a new and very promising class of high-performance and broadband THz emitters. Finally, this work is an example for a rapid translation of recently discovered fundamental physical effects into useful technology that can straightforwardly be employed by the broad femtosecond-laser community.

# Methods

**Samples: fabrication.** The magnetic heterostructures are grown on glass, sapphire or MgO substrates. Most of the samples from the Mainz group are fabricated using an Ar sputter deposition tool (Singulus Rotaris) with targets of 100 mm diameter. Typically, the Ar pressure range is 2 to $4 \times 10^{-3}$ mbar, the power used is 800 W, and deposition rates are 1.2 Å s$^{-1}$ for the ferromagnetic (FM) layer and 2.1 Å s$^{-1}$ for the nonferromagnetic (NM) layer. Prior to deposition, a short plasma etch is performed to clean the substrate surface from organic contaminants. The epitaxial Fe(100) thin film (thickness 3 nm) is prepared by RF-sputtering on a MgO(100) substrate at room temperature. After confirmation of the epitaxial growth by reflection high-energy electron diffraction, an epitaxial Pt(100) layer (3 nm) is DC-sputtered on top.

The $Co_{20}Fe_{60}B_{20}$ films from the Greifswald group are prepared by magnetron sputtering while NM metal films are grown by electron-beam evaporation under ultrahigh-vacuum conditions (base pressure $5 \times 10^{-10}$ mbar) using *in situ* transfer. Composition analysis of the films yields a Co/Fe ratio of 1/2.1, and characterization by transmission-electron microscopy reveals smooth $Co_{20}Fe_{60}B_{20}$ film surfaces below the atomic-monolayer limit.

The samples exhibit typical thin-film properties[53], in-plane magnetic anisotropy and a nearly rectangular hysteresis curve with a coercive field well below 10 mT.

**Samples: optical properties.** To further characterize our heterostructures, optical reflectance and transmittance are measured using the pump beam of our THz emission setup (see below). Our data (see Supplementary Fig. S3) show that the multilayers absorb about 50% of the incident laser power, largely independent of the total metal thickness. The measured values of reflectance and transmittance agree excellently with calculations based on a transfer-matrix formalism[54] and literature data of the optical constants of the materials involved[55,56]. Such good agreement is indicative of an optically homogeneous and flat metal film as expected from the optimized film deposition.

The THz conductivity of the Pt and FM films is measured by THz transmission spectroscopy[57]. We find values of $\sigma = (6.5 + 0.1i) \cdot 10^5$ S m$^{-1}$ for $Co_{40}Fe_{40}B_{20}$ and $(2.9 + 0.1i) \cdot 10^6$ S m$^{-1}$ for Pt, approximately independent of the THz frequency due to the high Drude scattering rate, which is on the order of 50 THz.

**Terahertz emission setup.** In the optical experiment (see Fig. 1a), the sample is kept in an external saturating magnetic field of 10 mT. It is excited by linearly polarized laser pulses (duration 10 fs, center wavelength 800 nm, energy 2.5 nJ) from a Ti:sapphire laser oscillator (repetition rate 80 MHz) under normal incidence from the substrate side (beam diameter at sample 50 μm full width at half maximum of the intensity). The THz electric field is detected by electrooptic sampling[1,2,47] where probe pulses (0.6 nJ, 10 fs) from the same laser copropagate with the THz field through an electrooptic crystal. The resulting signal $S(t)$ equals twice the THz-field-induced probe ellipticity where $t$ is the delay between the THz and probe pulse. Depending on signal strength, duration and bandwidth required, we use various electrooptic materials: ZnTe(110) (thickness of 10 μm and 1 mm)[47], GaP(110) (50 and 250 μm)[47] and the poled-polymer guest-host system Lemke/amorphous polycarbonate (LAPC)[14] (70 μm).

In the performance evaluation, the spintronic trilayer and the reference emitters (see Fig. 4a) are operated under identical pump-beam and detection conditions. The THz amplitude obtained from the commercially available photoconductive switch (TeraSED3, based on interdigitated electrodes on a semi-insulating GaAs substrate[22]) is maximized by setting the DC bias voltage to 12 V (20% above the maximum value recommended by the technical specifications). All measurements are performed at room temperature in a N$_2$ atmosphere.

**Signal deconvolution.** To extract the THz electric field $E_{\text{det}}$ incident onto the detector from the THz signal $S$ measured by electrooptic sampling, we note that these waveforms are connected by the convolution

$$S(t) = (h * E_{\text{det}})(t). \qquad (2)$$

The detector response function $h$ depends on the parameters of the electrooptic crystal and the sampling pulse used[47,58,59]. For the calculation of $h$, the optical constants are taken from Refs. 47, 60 and 61. Supplementary Figure S2 shows spectral amplitude and phase of the calculated $h(\omega)$ for the detectors used in this work. By equidistant sampling of the measured $S(t)$ and the calculated $h(t)$, we rewrite Eq. (2) as an overdetermined matrix equation and numerically solve for $E_{\text{det}}(t)$. Example traces of the absolute electric field are displayed in Supplementary Fig. S1a.

**Derivation of Eq. (1).** To derive a relationship between the emitted THz electric field and its source current, we note that within our sample the beam diameter is much larger than the sample thickness ($d \sim 10$ nm). Therefore, plane-wave propagation along the $z$-axis (see Fig. 1a) is assumed. The charge current density $-ej_c$ resulting from the laser-driven spin current density $(\hbar/2)j_s$ and the inverse spin Hall effect generates an electromagnetic wave with an electric field $E(z, t)$ polarized along the $x$-axis (see Fig. 1a). In the frequency domain, the dynamics of $E$ are governed by the wave equation[62]

$$[\partial_z^2 + k^2(z, \omega)]E(z, \omega) = Q(z, \omega) = -eZ_0 \omega\, j_c(z, \omega)/ic \qquad (3)$$

where $c$ is the vacuum speed of light, and $Z_0 = 377\,\Omega$ is the vacuum impedance. The THz wavevector $k(z, \omega)$ is given by $k^2 = k_0^2 + \Delta(k^2)$ where $k_0 = n\omega/c$ refers to the system without metal film for which the refractive index $n(z, \omega)$ equals $n_1(\omega)$ for the substrate half-space ($z < 0$) and $n_2 \approx 1$ for the air half-space ($z > 0$). When metal films are deposited on the substrate, the wavevector landscape $k_0^2(z, \omega)$ changes by $\Delta(k^2) = iZ_0 \sigma \omega/c$ where $\sigma(z, \omega)$ is the conductivity distribution of the metal stack. We omit the $\omega$-dependence in the notation for brevity and rewrite Eq. (3) as an integral equation

$$E(z) = \int dz' [Q(z') - \Delta(k^2)(z')E(z')]\, G_0(z, z') \qquad (4)$$

where $G_0(z, z')$ is the Green's function of the system without metal films ($k = k_0$)[62]. When both source point $z'$ and observation point $z$ are on the air side ($z, z' > 0$), one has $G_0(z, z') = (e^{ik_2|z-z'|}/2ik_2) \cdot (1 + r_{21}e^{2ik_2|z|})$ where $k_2 = n_2\omega/c$. The term with the Fresnel coefficient $r_{21} = (n_2 - n_1)/(n_2 + n_1)$ accounts for wave reflection at the air-substrate interface at $z = 0$.

Since the film is much thinner than the wavelength and attenuation length of the THz wave, we approximate all phase factors by 1 and assume the electric field $E$ is constant throughout the metal multilayer (quasistatic approximation). As a consequence, both $E$ and $G_0 \approx (c/i\omega)/(n_1 + n_2)$ can be moved in front of the integral of Eq. (4), and we obtain Eq. (1) of the main text in which the impedance is given by

$$\frac{1}{Z(\omega)} = \frac{n_1(\omega) + n_2(\omega)}{Z_0} + \int_0^d dz\, \sigma(z, \omega). \qquad (5)$$

When the sheet conductance $\int dz\, \sigma$ of the metal stack is much larger than the shunt conductance $(n_1 + n_2)/Z_0$ of the two adjacent half-spaces, Eq. (1) turns into the familiar Ohm's law[44]. Note that our derivation accounts for the propagation of the THz radiation inside the sample (including all reflection echoes) and the irradiation into free space (see Fig. 3b). It is, however, only valid in the thin-film limit.

**Model for metal-thickness dependence of THz emission.** To model the THz emission amplitude of a FM/NM bilayer, we make use of Eqs. (1) and (5) and assume that FM=$Co_{20}Fe_{60}B_{20}$ and NM=Pt. The impedance of the bilayer [Eq. (5)] is determined by the measured THz conductivities of the materials involved (see above).

Equation (1) requires knowledge of the charge current density $j_c(z) = \gamma(z)j_s(z)$. To model its spatial structure, we neglect the spin-Hall angle $\gamma$ in the FM layer and determine the spin current density $j_s$ in the NM layer by considering the following simplified scenario: after excitation by the pump pulse, spin-polarized hot electrons from the FM layer enter the NM layer in which they first propagate ballistically away from the FM/NM interface (see Fig. 1a). However, once an electron undergoes scattering, its velocity is randomized such that its contribution to the ultrafast photocurrent and, thus, THz signal becomes negligible. To account for such velocity relaxation, we assume that the density of ballistic electrons behind the FM/NM interface decreases according to $e^{-(z-d_{\text{FM}})/\lambda_{\text{rel}}}$. Here, $\lambda_{\text{rel}}$ can be considered as a hot-electron velocity relaxation length, and $d_{FM}$ is the thickness of the FM layer. We furthermore assume that the electrons undergo perfect reflection at the NM/air and the NM/FM interface.

As shown in Ref. 63, these assumptions imply a spatial dependence of the ballistic spin current density inside the NM layer ($d_{FM} < z < d = d_{FM} + d_{NM}$) according to

$$j_s(z) = j_s(d_{FM}) \frac{\sinh[(z-d)/\lambda_{rel}]}{\sinh(d_{NM}/\lambda_{rel})} \qquad (6)$$

where $j_s(d_{FM})$ is the spin current density directly after the FM layer. Finally, the linear fluence dependence of the emitted THz field (see Fig. 1b) indicates that the spin current is proportional to the energy density deposited by the pump pulse. Therefore, one has $j_s(d_{FM}) \propto A/d$ where $A$ is the absorbed fraction of the incident pump power (see Supplementary Fig. S3). Plugging Eq. (6) and all other assumptions into Eqs. (1) and (5), we obtain

$$E(d) \propto \gamma_{NM} \lambda_{rel} \frac{A}{d} \frac{\tanh(d_{NM}/2\lambda_{rel})}{n_1 + n_2 + Z_0 \int_0^d dz\ \sigma(z)} \qquad (7)$$

with $\lambda_{rel}$ and a global amplitude factor being the only free parameters.

The choice $\lambda_{rel}$=1 nm yields the best fit to the measured thickness dependence of the emitted THz amplitude, including the existence of a maximum (see Supplementary Fig. S6). The discrepancy for $d$<4 nm (Fig. 3a) is most likely due to shortcomings of our simplified transport model and due to changes in the magnetic properties (such as the Curie temperature) of the sample when approaching small thicknesses $d$. For $d$>25 nm, we do not expect additional Fabry-Pérot resonances for the pump beam because the decay length of the pump intensity (~15 nm) is much shorter than the wavelength (~300 nm) of the pump radiation inside the metal film.

*Ab initio* **calculations of the spin-Hall effect.** The spin-Hall conductivity (see Fig. 2) is calculated by means of the Kubo formula within density-functional theory using the full-potential linearized augmented plane-wave (FLAPW) program FLEUR (see http://www.flapw.de). We employ the generalized gradient approximation of the exchange correlation potential, a plane-wave cutoff at a wavevector of 85 nm$^{-1}$ and the experimental lattice constants (2.91, 3.892, 3.302, 3.166, 3.8402 and 3.926 Å for Cr, Pd, Ta, W, Ir and Pt, respectively). Further details on the computation are given in Ref. 64.

## Acknowledgements

We are grateful to S. Winnerl for providing us with a TeraSED3 emitter and for stimulating discussions. The authors thank the German Science Foundation for funding through SPP 1538/SpinCaT (Berlin, Greifswald, Jülich and Mainz groups) and through SFB TRR 173/Spin+X (Mainz group). The European Union is gratefully acknowledged for funding through the ERC H2020 CoG project TERAMAG/grant no. 681917 (T.K.), the FP7 project CRONOS/grant no. 280879 (T.K. and M.W.), the FP7 Marie Curie ITN WALL project/grant no. 608031 (J.S., G.J. and M.K.), the Career Integration Grant LIGHTER/grant no. 334324 (D.T.) and the FP7 projects Fantomas/grant no. 214810 and FemtoSpin/grant no. 281043 (P.M. and P.M.O.). The authors are grateful for support by the Max Planck Society (T.K. and D.T.), the Swedish Research Council (P.M. and P.M.O.), the Samsung SGMI program, the EFRE Program of the state of Rhineland Palatinate, TT-DINEMA, the Excellence Graduate School MAINZ (GSC 266) and the state research centre CINEMA (Mainz group). F.F. and Y.M. are grateful for computing time at Jülich Supercomputing Centre.

## Author contributions

T.K. and T.S. conceived the experiments. S.J., U.M., A.K., J.He., E.B., M.J., G.J., M.M. and M.K. fabricated the spintronic emitters and optimized the fabrication process. L.B., T.S. and T.K. built the THz emission setup. J.Ha. and L.M.H. fabricated the LAPC electrooptic detectors. T.S. performed the THz experiments and optical sample characterization. T.S. and T.K. analyzed experimental data with contributions from D.T. and M.K. T.K. and T.S. developed the analytical THz emitter model with contributions from F.F. and Y.M. F.F. and Y.M. calculated spin-Hall conductivities. P.M. and P.M.O. conducted supporting calculations of the ultrafast spin transport. T.K., T.S., D.T., M.W. and M.K. co-wrote the paper. All authors contributed to discussions of the results and to writing the paper.

## Additional information

Correspondence and requests for materials should be addressed to T.K.

# Reference

[1] Ferguson, B. & Zhang, X. Materials for terahertz science and technology. *Nature Materials* **1**, 26-33 (2002).

[2] Tonouchi, M. Cutting-edge terahertz technology. *Nature Photon.* **1**, 97-105 (2007).

[3] Ulbricht, R., Hendry, E., Shan, J., Heinz, T. & Bonn, M. Carrier dynamics in semiconductors studied with time-resolved terahertz spectroscopy. *Rev. Mod. Phys.* **83**, 543-586 (2011).

[4] Kampfrath, T., Tanaka, K. & Nelson, K. Resonant and nonresonant control over matter and light by intense terahertz transients. *Nature Photon.* **7**, 680-690 (2013).

[5] Chan, W., Deibel, J. & Mittleman, D. Imaging with terahertz radiation. *Rep. Prog. Phys.* **70**, 1325-1379 (2007).

[6] Cocker, T.L. *et al.* An ultrafast terahertz scanning tunnelling microscope. *Nature Photon.* **7**, 620-625 (2013).

[7] Sano, Y. *et al.* Imaging molecular adsorption and desorption dynamics on graphene using terahertz emission spectroscopy. *Sci. Rep.* **4**, 6046; DOI:10.1038/srep06046 (2014).

[8] Zeitler, J. *et al.* Terahertz pulsed spectroscopy and imaging in the pharmaceutical setting - a review. *Journal of Pharmacy and Pharmacology* **59**, 209-223 (2007).

[9] Reimann, K. Table-top sources of ultrashort THz pulses. *Rep. Prog. Phys.* **70**, 1597-1632 (2007).

[10] Blanchard, F. *et al.* Generation of Intense Terahertz Radiation via Optical Methods. *IEEE Journal of Selected Topics in Quantum Electronics* **17**, 5-16 (2011).

[11] Rice, A. *et al.* Terahertz optical rectification from ⟨110⟩ zinc-blende crystals. *Appl. Phys. Lett.* **64**, 1324-1326 (1994).

[12] Kaindl, R., Eickemeyer, F., Woerner, M. & Elsaesser, T. Broadband phase-matched difference frequency mixing of femtosecond pulses in GaSe: Experiment and theory. *Appl. Phys. Lett.* **75**, 1060-1062 (1999).

[13] Huber, R., Brodschelm, A., Tauser, F. & Leitenstorfer, A. Generation and field-resolved detection of femtosecond electromagnetic pulses tunable up to 41 THz. *Appl. Phys. Lett.* **76**, 3191-3193 (2000).

[14] Zheng, X., Sinyukov, A. & Hayden, L.M. Broadband and gap-free response of a terahertz system based on a poled polymer emitter-sensor pair. *Appl. Phys. Lett.* **87**, 081115 (2005).

[15] Zheng, X., McLaughlin, C.V., Cunningham, P., & Hayden, L.M. Organic broadband terahertz sources and sensors. *Journal of Nanoelectronics and Optoelectronics* **2**, 58-76 (2007).

[16] Brunner, F. et al. A hydrogen-bonded organic nonlinear optical crystal for high-efficiency terahertz generation and detection. *Opt. Express* 16, 16496-16508 (2008).

[17] Katayama, I., Akai, R., Bito, M., Shimosato, H., Miyamoto, K., Ito, H. & Ashida, M. Ultrabroadband terahertz generation using 4-N,N-dimethylamino-4'-N'-methyl-stilbazolium tosylate single crystals. *Appl. Phys. Lett.* **97**, 021105 (2010).

[18] Shan, J., & Heinz, T.F. Terahertz radiation from semiconductors. In *Ultrafast Dynamical Processes in Semiconductors* (pp. 1-56). Springer Berlin Heidelberg (2004).

[19] Apostolopoulos, V. & Barnes, M. THz emitters based on the photo-Dember effect. *Journal of Physics D: Applied Physics* **47**, 374002 (2014).

[20] Klatt, G. *et al.* Terahertz emission from lateral photo-Dember currents. *Opt. Express* **18**, 4939-4947 (2010).

[21] Shen, Y., Upadhya, P., Linfield, E., Beere, H. & Davies, A. Ultrabroadband terahertz radiation from low-temperature-grown GaAs photoconductive emitters. *Appl. Phys. Lett.* **83**, 3117-3119 (2003).

[22] Winnerl, S. Scalable Microstructured Photoconductive Terahertz Emitters. *Journal of Infrared, Millimeter, and Terahertz Waves* **33**, 431-454 (2012).

[23] Hale, P. *et al.* 20 THz broadband generation using semi-insulating GaAs interdigitated photoconductive antennas. *Opt. Express* **22**, 26358-26364 (2014).

[24] Berry, C., Wang, N., Hashemi, M., Unlu, M. & Jarrahi, M. Significant performance enhancement in photoconductive terahertz optoelectronics by incorporating plasmonic contact electrodes. *Nature Communications* **4**, 1622 (2013).

[25] Thomson, M.D., Kreß, M., Löffler, T. & Roskos, H.G. Broadband THz emission from gas plasmas induced by femtosecond optical pulses: From fundamentals to applications. *Laser & Photon. Rev.* **1**, 349–368 (2007).

[26] Kim, K., Glownia, J., Taylor, A.J. & Rodriguez, G. High-Power Broadband Terahertz Generation via Two-Color Photoionization in Gases. *IEEE Journal of Quantum Electronics* **48**, 797-805 (2012).

[27] Lu, X. & Zhang, X.C. Investigation of ultra-broadband terahertz time-domain spectroscopy with terahertz wave gas photonics. *Frontiers of Optoelectronics* **7**, 121-155 (2013).

[28] Bartel, T., Gaal, P., Reimann, K., Woerner, M. & Elsaesser, T. Generation of single-cycle THz transients with high electric-field amplitudes. *Opt. Lett.* **30**, 2805-2807 (2005).

[29] Buccheri, F. & Zhang, X.C. Terahertz emission from laser-induced microplasma in ambient air. *Optica* **2**, 366-369 (2015).

[30] Ramanandan, G., Ramakrishnan, G., Kumar, N., Adam, A. & Planken, P. Emission of terahertz pulses from nanostructured metal surfaces. *Journal of Physics D: Applied Physics* **47**, 374003 (2014).

[31] Zhang, L. et al. High-power THz to IR emission by femtosecond laser irradiation of random 2D metallic nanostructures. *Sci. Rep.* **5**, 12536 (2015).

[32] Zhukov, V., Chulkov, E. & Echenique, P. Lifetimes and inelastic mean free path of low-energy excited electrons in Fe, Ni, Pt, and Au: Ab initio GW + T calculations. *Phys. Rev. B* **73**, 125105 (2006).

[33] Laman, N. & Grischkowsky, D. Terahertz conductivity of thin metal films. *Appl. Phys. Lett.* **93**, 051105 (2008).

[34] Ramakrishnan, G. & Planken, P. Percolation-enhanced generation of terahertz pulses by optical rectification on ultrathin gold films. *Optics Letters* **36**, 2572-2574 (2011).

[35] Polyushkin, D., Hendry, E., Stone, E. & Barnes, W. THz Generation from Plasmonic Nanoparticle Arrays. *Nano Letters* **11**, 4718-4724 (2011).

[36] Kadlec, F., Kuzel, P. & Coutaz, J. Study of terahertz radiation generated by optical rectification on thin gold films. *Optics Letters* **30**, 1402-1404 (2005).

[37] Welsh, G.H., Hunt, N.T. & Wynne, K. Terahertz-Pulse Emission Through Laser Excitation of Surface Plasmons in a Metal Grating. *Phys. Rev. Lett.* **98**, 026803 (2007).

[38] Saitoh, E., Ueda, M., Miyajima, H. & Tatara, G. Conversion of spin current into charge current at room temperature: Inverse spin-Hall effect. *Appl. Phys. Lett.* **88**, 182509 (2006).

[39] Hoffmann, A. Spin Hall Effects in Metals. *IEEE Trans. Magn.* **49**, 5172-5193 (2013).

[40] Sinova, J. *et al.* Spin Hall Effects. *Reviews of Modern Physics* **87**, 1213-1260 (2015).

[41] Wei, D. *et al.* Spin Hall voltages from a.c. and d.c. spin currents. *Nature Communications* 5:3768 DOI: 10.1038/ncomms4768 (2014).

[42] Battiato, M., Carva, K. & Oppeneer, P.M. Superdiffusive spin transport as a mechanism of ultrafast demagnetization. *Phys. Rev. Lett.* **105**, 027203 (2010).

[43] Melnikov, A. *et al.*, Ultrafast transport of laser-excited spin polarized carriers in Au/Fe/MgO(001). *Phys. Rev. Lett.* **107**, 076601 (2011).

[44] Kampfrath, T. *et al.* Terahertz spin current pulses controlled by magnetic heterostructures. *Nature Nanotech* **8**, 256-260 (2013).

[45] Héroux, J.B., Ino, Y., Kuwata-Gonokami, M., Hashimoto, Y. & S. Katsumoto, S. Terahertz radiation emission from GaMnAs. *Appl. Phys. Lett.* **88**, 221110 (2006).

[46] Jin, Z. *et al.* Accessing the fundamentals of magnetotransport in metals with terahertz probes. *Nature Physics* **11**, 761-766 (2015).

[47] Leitenstorfer, A., Hunsche, S., Shah, J., Nuss, M. & Knox, W. Detectors and sources for ultrabroadband electro-optic sampling: Experiment and theory. *Appl. Phys. Lett.* **74**, 1516-1518 (1999).

[48] Kontani, H., Tanaka, T., Hirashima, D., Yamada, K. & Inoue, J. Giant Orbital Hall Effect in Transition Metals: Origin of Large Spin and Anomalous Hall Effects. *Phys. Rev. Lett.* **102**, 016601 (2009).

[49] D'Angelo, F., Mics, Z., Bonn, M. & Turchinovich, D. Ultra-broadband THz time-domain spectroscopy of common polymers using THz air photonics. *Opt. Express* **22**, 12475-12485 (2014).

[50] Blanchard, F. et al. Generation of 1.5 µJ single-cycle terahertz pulses by optical rectification from a large aperture ZnTe crystal. *Opt. Express* **15**, 13212–13220 (2007).

[51] Liu, M. et al. Terahertz-field-induced insulator-to-metal transition in vanadium dioxide metamaterial. *Nature* **487**, 345–348 (2012).

[52] Vicario, C. et al. Off-resonant magnetization dynamics phase-locked to an intense phase-stable terahertz transient. *Nat. Photonics* **7**, 720–723 (2013).

**--- Refs of Methods section ---**

[53] Boulle, O., Malinowski, G. & Kläui, M. Current-induced domain wall motion in nanoscale ferromagnetic elements. *Materials Science and Engineering: R: Reports* **72**, 159-187 (2011).

[54] Yeh, P. *Optical waves in layered media* (Wiley, 2005).

[55] Liang, X., Xu, X., Zheng, R., Lum, Z. & Qiu, J. Optical constant of CoFeB thin film measured with the interference enhancement method. *Appl. Opt.* **54**, 1557-1563 (2015).

[56] Ordal, M., Bell, R., Alexander, Jr., R., Newquist, L. & Querry, M. Optical properties of Al, Fe, Ti, Ta, W, and Mo at submillimeter wavelengths. *Appl. Opt.* **27**, 1203-1209 (1988).

[57] Nuss, M.C. & Orenstein, J. in Millimeter and Submillimeter Wave Spectroscopy of Solids (ed. Gruener, G.) Ch. 2 (Springer, Berlin, 1998).

[58] Kampfrath, T., Nötzold, J. & Wolf, M. Sampling of broadband terahertz pulses with thick electrooptic crystals. *Appl. Phys. Lett.* **90**, 231113 (2007).

[59] Gallot, G. & Grischkowsky, D. Electro-optic detection of terahertz radiation. *Journal of the Optical Society of America B* **16**, 1204-1212 (1999).

[60] Dietze, D., Unterrainer, K. & Darmo, J., Dynamically phase-matched terahertz generation. *Opt. Lett.* **37**, 1047-1049 (2012).

[61] Zheng, X., McLaughlin, C., Leahy-Hoppa, M., Sinyukov, A. & Hayden, L. Modeling a broadband terahertz system based on an electro-optic polymer emitter-sensor pair. *Journal of the Optical Society of America B* **23**, 1338-1347 (2006).

[62] Mills, D.L. *Nonlinear optics: Basic concepts* (Springer, 1991).

[63] Mosendz, O. et al. Quantifying Spin Hall Angles from Spin Pumping: Experiments and Theory. *Phys. Rev. Lett.* **104**, 046601 (2010).

[64] Freimuth, F., Blügel, S. & Mokrousov, Y. Anisotropic spin Hall effect from first principles. *Phys. Rev. Lett.* **105**, 246602 (2010).

# Efficient metallic spintronic emitters of ultrabroadband terahertz radiation: supplementary information

T. Seifert, S. Jaiswal, U. Martens, J. Hannegan, L. Braun, P. Maldonado, F. Freimuth, A. Kronenberg, J. Henrizi, I. Radu, E. Beaurepaire, Y. Mokrousov, P.M. Oppeneer, M. Jourdan, G. Jakob, D. Turchinovich, L.M. Hayden, M. Wolf, M. Münzenberg, M. Kläui, T. Kampfrath

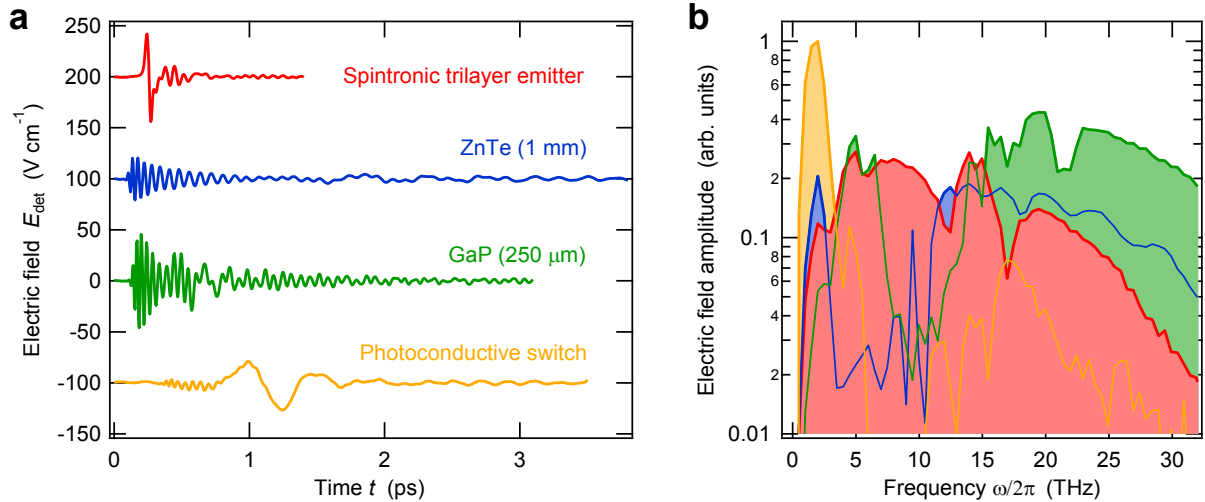

**Figure S1 | THz electric field emitted by the spintronic source and other emitters. a**, Transient THz electric field $E_{det}$ incident on the electrooptic detection crystal following emission by four THz emitters: the spintronic trilayer emitter W(2 nm)/Co$_{40}$Fe$_{40}$B$_{20}$(1.8 nm)/Pt(2 nm), the (110)-oriented nonlinear-optical crystals ZnTe (thickness 1 mm) and GaP (250 µm), and a photoconductive switch. **b**, Corresponding field amplitudes obtained by Fourier transformation of the data of panel **a**. All emitters are driven by pump pulses (duration 10 fs, energy 2.5 nJ, centre wavelength 800 nm) from a Ti:sapphire laser oscillator (repetition rate 80 MHz). The THz signals are measured with a 1 mm thick ZnTe(110) electrooptic crystal. See the main text and Methods for more details.

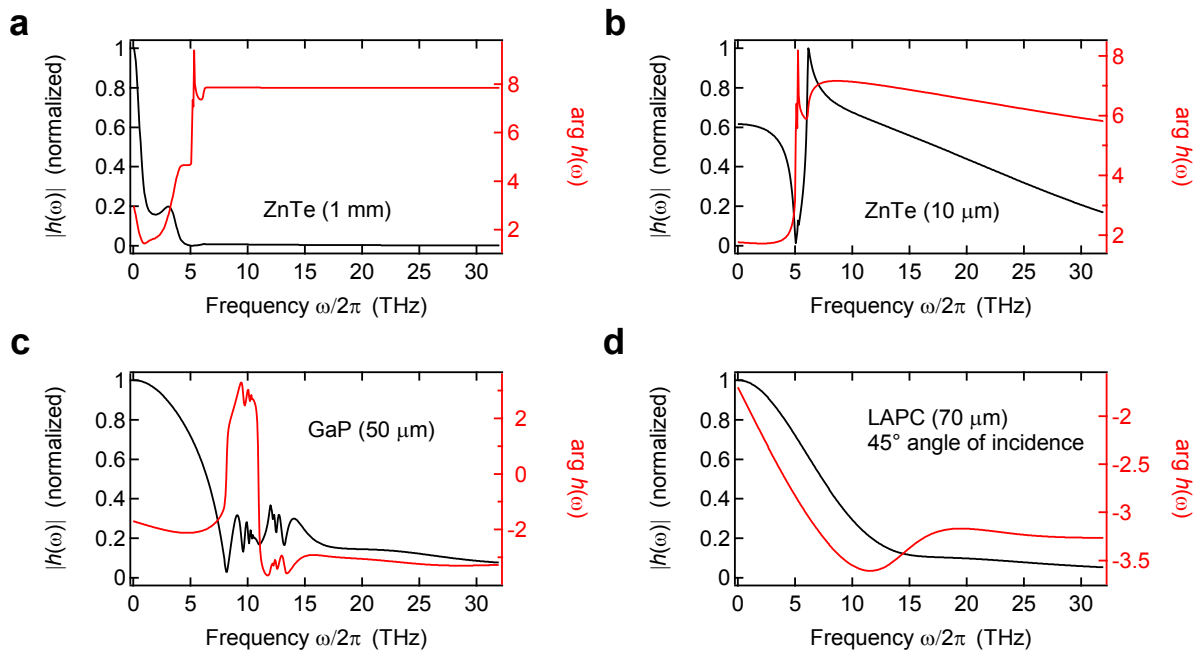

**Figure S2 | Detector response.** Spectral amplitude and phase of the calculated response functions $h(\omega)$ of the electrooptic detectors used in this work (see Methods): **a**, 10 µm ZnTe(110) on an inactive ZnTe(001) substrate, **b**, 1 mm ZnTe(110), **c**, 50 µm GaP(110) and **d**, 70 µm Lemke/amorphous polycarbonate (LAPC) under 45° angle of incidence. The amplitude maxima are normalized to 1.

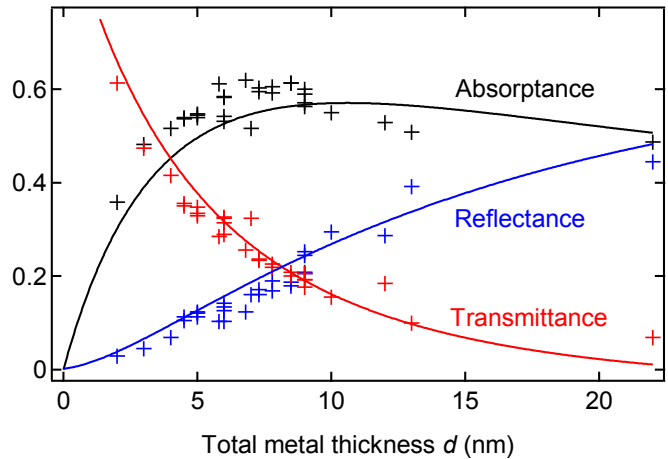

**Figure S3 | Optical properties of the metal heterostructures.** Reflectance, transmittance and resulting absorptance $A$ of metal stacks as a function of metal thickness $d$, measured by using the optical pump beam of the THz emission setup. The solid lines show calculations (see Methods).

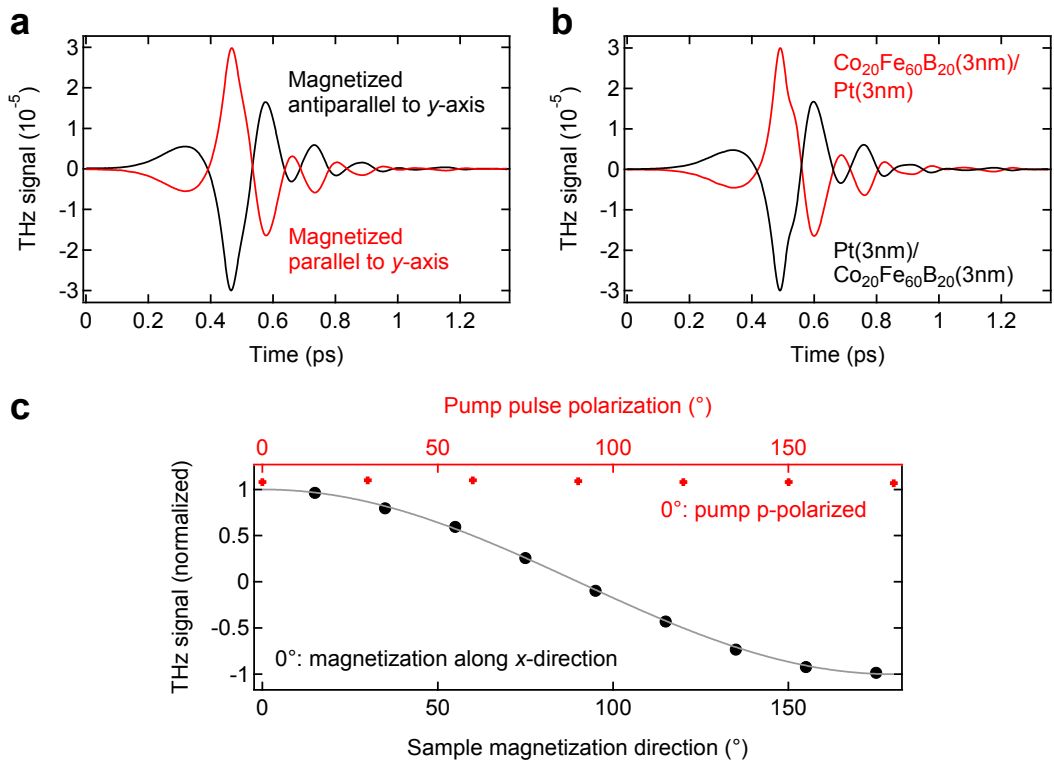

**Figure S4 | Symmetry properties of the emitted THz radiation. a**, THz signal for the FM layer magnetization parallel and antiparallel to the $y$-axis (see Fig. 1a main text). **b**, THz signal obtained for an FM/NM bilayer and its NM/FM counterpart having reversed layer stacking (see labels for sample details). **c**, THz signal amplitude as a function of the orientation of the linear pump polarization (red crosses) and as a function of the direction of the sample magnetization (black circles). For the latter measurement, a THz wire-grid polarizer was placed before the detection to block the $x$-polarized THz field component (see Fig. 1a main text). The grey solid line is the cosine of the angle between the polarizer wires and the sample magnetization.

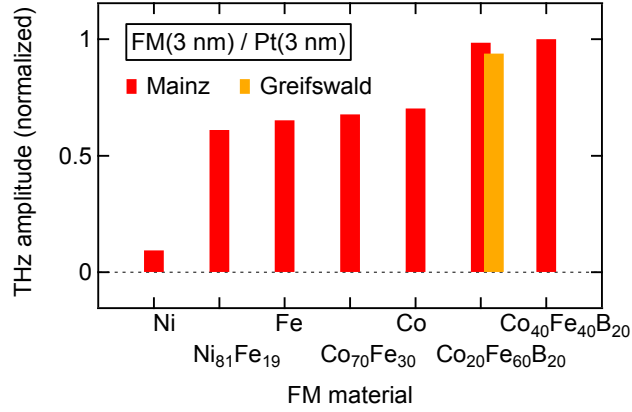

**Figure S5 | Impact of the FM material.** THz signal amplitude (RMS) of a FM(3 nm)/Pt(3 nm) heterostructure as a function of the FM material chosen. Different colours indicate different labs for sample fabrication (see Methods).

***S1. Variation of the FM material.*** To study the impact of the FM material on THz emission, we consider FM(3 nm)/Pt(3 nm) heterostructures with the FM metals Fe, Co, Ni and their binary alloys. Results are displayed in Fig. 2b and show that all materials provide similar THz output, except Ni, which yields less than 20% of the maximum THz amplitude. The reason for this behaviour is not yet understood but may be related to the fact that the Curie temperature of Ni (627 K) is considerably lower than that of all other FM materials (>1000 K).

***S2. Estimate of the quadratic nonlinear susceptibility.*** The amplitude of the THz pulses emitted from the spintronic THz emitter (STE) grows linearly with the pump fluence (Fig. 1b main text). Therefore, optically induced THz emission from the STE can be described phenomenologically as a $\chi^{(2)}$-type nonlinear-optical process in which the pump field $E_\mathrm{p}$ induces an electric dipole density (polarization)[65]

$$P^{(2)}(\omega) = \frac{\chi^{(2)}(\omega)}{Z_0 c} \int_{\omega_\mathrm{p}>\omega>0} \mathrm{d}\omega_\mathrm{p}\, E_\mathrm{p}(\omega_\mathrm{p}) E_\mathrm{p}^*(\omega_\mathrm{p} - \omega) = \frac{\chi^{(2)}(\omega)}{Z_0 c} F(\omega) \qquad (S8)$$

at the difference frequency $\omega$ of all pump-frequency pairs $\omega_\mathrm{p}$ and $\omega_\mathrm{p} - \omega$. In writing Eq. (S8), we have assumed that the quadratic response function $\chi^{(2)}$ is independent of the pump frequency and abbreviated the integral by $F(\omega)$. In general, the dependence of $\chi^{(2)}$ on THz frequency $\omega$ makes the response non-instantaneous, that is, longer than the driving optical pump pulse. Since $-\mathrm{i}\omega P^{(2)}(\omega)$ equals the charge-current density $-ej_\mathrm{c}(\omega)$, we use Eqs. (1) and (5) of the main text to rewrite the THz electric field directly after the STE as

$$E(\omega) = \frac{1}{n_1(\omega) + n_2(\omega) + Z_0 \int_0^d dz\, \sigma(\omega)} \cdot \frac{i\omega F(\omega)}{c} \cdot \int_0^d dz\, \chi^{(2)}(\omega). \quad (S9)$$

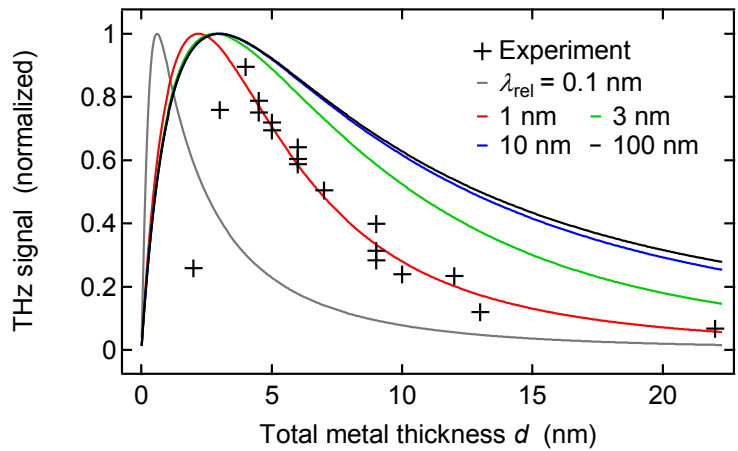

**Figure S6 | Thickness dependence of THz emission.** Normalized measured (crosses) and modelled (solid lines) THz signal amplitude as a function of total metal thickness $d$. Calculations are shown for 5 different values of the hot-electron velocity relaxation length $\lambda_\mathrm{rel}$.

To estimate the $\chi^{(2)}$ magnitude of our STE, we compare the THz amplitude generated by the STE to that of a suitable reference emitter. We choose a 50 μm thick GaP(110) crystal because the magnitude ($5.4\times10^{-11}$ m V$^{-1}$) of its only nonvanishing $\chi^{(2)}$ tensor element is well known[66]. In addition, THz-wave generation is still phase-matched at this thickness. According to Ref. 67, the THz amplitude directly after the GaP is given by a formula analogous to Eq. (S9),

$$E_{\text{GaP}}(\omega) = \frac{\exp[in_{\text{GaP}}(\omega)\omega d/c]}{n_2(\omega) + n_{\text{GaP}}(\omega)} \cdot \frac{i\omega F_{\text{GaP}}(\omega)}{c} \cdot d_{\text{GaP}} \chi^{(2)}_{\text{GaP}}(\omega). \quad (S10)$$

Here, $d_{\text{GaP}}$ is the GaP thickness, and $n_{\text{GaP}}$ is the refractive index of GaP. As with Eq. (S9), $n_2$ is the refractive index of air.

In our experiment, we find that the peak electrooptic signal (Fig. 4a main text) and the THz field amplitude (Fig. S1a) of the STE are comparable to those of GaP. Consequently, we have $|E|\sim|E_{\text{GaP}}|$. Similarly, the first and second term of Eq. (S9) have the same order of magnitude as their GaP counterparts in Eq. (S10). Therefore, the ratio of the nonlinear-optical coefficients is roughly given by

$$\left|\chi^{(2)}/\chi^{(2)}_{\text{GaP}}\right| \sim d_{\text{GaP}}/d_{\text{eff}}, \quad (S11)$$

that is, by the ratio of the GaP thickness ($d_{\text{GaP}} = 50$ μm) and the effective thickness $d_{\text{eff}}$ of the STE region in which the THz charge current is generated. Since $d_{\text{eff}}\sim\lambda_{\text{rel}}\sim 1$ nm (see main text and Methods), the $\chi^{(2)}$ coefficient of the STE is nearly 5 orders of magnitude larger than that of GaP, yet located in a sheet of only ~1 nm thickness at the FM/NM interface.

# References

[65] Mills, D.L. *Nonlinear optics: Basic concepts* (Springer, 1991).
[66] Leitenstorfer, A., Hunsche, S., Shah, J., Nuss, M. & Knox, W. Detectors and sources for ultrabroadband electro-optic sampling: Experiment and theory. *Appl. Phys. Lett.* **74**, 1516-1518 (1999).
[67] Kaindl, R.A. *et al.* Broadband phase-matched difference frequency mixing of femtosecond pulses in GaSe: Experiment and theory. *Appl. Phys. Lett.* **75**, 1060-1062 (1999).